\documentclass[11pt]{article}
\usepackage{pstricks}
\usepackage{pst-plot}
\renewcommand{\arraystretch}{0.9}

\font\eufm=eufm10
\def\frak#1{\hbox{\eufm#1}}
\newcommand{\bd}{
\begin{document}}
\newcommand{\ed}{\end{document}}
\newcommand{\be}{\begin{enumerate}}
\newcommand{\ee}{\end{enumerate}}
\newcommand{\bi}{\begin{itemize}}
\newcommand{\ei}{\end{itemize}}
\newcommand{\ba}{\begin{array}}
\newcommand{\ea}{\end{array}}
\newcommand{\vs}{\vspace*{0.3\baselineskip}}
\newcommand{\vsm}{\vspace*{-0.3\baselineskip}}
\newcommand{\kom}[1]{}
\newtheorem{defi}{Definition}[section]
\newtheorem{tw}[defi]{Theorem}
\newtheorem{prop}[defi]{Proposition}
\newtheorem{lem}[defi]{Lemma}
\newtheorem{re}[defi]{Remark}
\newtheorem{col}[defi]{Corollary}
\newtheorem{ex}[defi]{Examples}
\newtheorem{zad}{Exercise}[section]
%
\newcommand{\Om}{\Omega}
\newcommand{\om}{\omega}
\newcommand{\G}{\Gamma}
\newcommand{\D}{\Delta}
\renewcommand{\d}{\delta}
\newcommand{\ga}{\gamma}
\newcommand{\eps}{\epsilon}
\newcommand{\C}{C}
\newcommand{\R}{R}
%
\newcommand{\ove}{\overline}
\newcommand{\ms}{\oplus}
\newcommand{\mt}{\otimes}
\newcommand{\dz}{\wedge}
\newcommand{\lra}{\longrightarrow}
\newcommand{\rel}{\mbox{$\,$\rule[0.5ex]{1.1em}{0.2pt}$\triangleright\,$}}
\newcommand{\dow}{\hspace*{\fill}\rule{1.6ex}{1.6ex}\hspace*{1em}}
\newcommand{\dowl}{\hspace*{\fill}\rule{1ex}{1ex}\hspace*{1em}}
\newcommand{\sd}{\hspace{0.3ex}\tiny{\rhd\mbox{\hspace{-2ex}}<}\hspace{0.3ex}}
%
\newcommand{\g}{\frak g}
\newcommand{\ab}{\frak a}
\newcommand{\bb}{\frak b}
\newcommand{\h}{\frak h}
\newcommand{\got}{\frak t}
\newcommand{\hd}{\hat{\d}}
\newcommand{\oml}{\Omega_L^{1/2}}
\newcommand{\omr}{\Omega_R^{1/2}}
\newcommand{\omh}{\Omega^{1/2}}
\newcommand{\lo}{\lambda_0}
\newcommand{\ro}{\rho_0}
\newcommand{\sA}{\mbox{$\cal A\,$}}
\newcommand{\sT}{\mbox{$\cal T\,$}}
\newcommand{\sB}{\mbox{$\cal B\,$}}
\newcommand{\lma}{\Lambda^{max}}
\newcommand{\timh}{\times_h}
\newcommand{\Gd}{\G^{(2)}}
\newcommand{\el}{e_L}
\newcommand{\er}{e_R}
\newcommand{\GG}{\G_1\times\G_2}
\newcommand{\gdot}{\hspace{-0.1em}\cdot\hspace{-0.1em}}
\newcommand{\tran}{\frown\hspace{-2.2ex}|\hspace{1.9ex}}
%
\newcommand{\la}[2]{\Lambda_{#1#2}}
\newcommand{\kad}{ad^{\#}}
\newcommand{\wl}[1]{\vphantom{X}_{#1}{\G}}
\newcommand{\te}{\tilde{e}}
\newcommand{\sD}{\mbox{$\cal D\,$}}
\newcommand{\notka}[1]{}
\newcommand{\dif}{differential }
\newcommand{\gru}{groupoid }
\newcommand{\grus}{groupoids }
\newcommand{\ti}{\tilde}
\newcommand{\halden}{half density }
\newcommand{\haldens}{half densities }
\renewcommand{\top}{topological }
%
\title{Towards a topological (dual of) quantum $\kappa$-Poincar\'{e} group}

\date{}
\author{Piotr Stachura\\
Department of Mathematical Methods in Physics,\\
University of Warsaw,  Poland
\thanks{The work was supported by Polish KBN grant No. 2 P03A 040 22}\\
e-mail: stachura@fuw.edu.pl}
\bd
\maketitle
\noindent{\bf Abstract\hspace{1em}} {\small We argue that the $\kappa$-deformation is related
to a factorization of a Lie group, therefore {\em an approproate version of $\kappa$-Poincar\'{e} 
does exist on the $C^*$-algebraic level}. The explict form of this factorization is computed 
that leads to an ``action'' of the Lorentz  group (with space reflections) 
considered in Doubly Special Relativity theory. The orbit structure is found and ``the
momentum manifold'' is extended in a way that removes singularities of the ``action'' and results
in a true action. Some global properties of this manifold are investigated.} 
\section{Introduction}

In  recent years there has been  some interest in relativistic theories with two
invariants: the speed of light $c$ and the invariant length $\lambda$,
usually identified more or less with the Planck length.  These are
``Doubly Special Relativity'' (DSR) theories. I do not  feel competent enough to 
discuss quantum  gravity arguments   and  experimental  data  in favour or 
against DSR and  refer to \cite{AC3,KG}  and references
therein for an overview of the subject. The first example of a DSR theory was proposed
by G. Amelino-Camelia in \cite{AC1,AC2} and now is usually called DSR1 theory. 
The main features of  the DSR1 is  a modified relation between energy and momentum 
\begin{equation}\label{dispersion}
\cosh( \lambda E) -\frac{\lambda^2}{2} p^2 e^{\lambda E}=\cosh(\lambda m)
\end{equation}
and a non linear action of the Lorentz group on the ``momentum space'' $(E,p)$ that
preserves this relation. On the other hand the relation (\ref{dispersion}) 
had been known long before  DSR1 framework appeared, 
namely as an expression for the Casimir in the $\kappa$-Poincar\'{e}  
algebra \cite{Luk, Maj}. There are different views on the issue
whether ``DSR1 framework'' and ``$\kappa$-Poincar\'{e} framework'' are equivalent,  
and what is the physical content of the DSR1 theory. I will not discuss them here  
and simply adopt the position that some version of $\kappa$-Poincar\'{e} is necessary 
for DSR1. The question is ``What is a $\kappa$-Poincar\'{e} framework ?'' The analysis
that follows contains mostly known facts nevertheless it also tries to clarify some points. 
It is intensionally informal  but most of statements can be made precise and rigorous using 
the language of $C^*$-algebras and methods presented in \cite{PS1,PS2}.

Consider an algebra generated by symbols $(N_i,M_j, P_\mu)\,\,i,j=1,2,3,\,\mu=0,1,2,3$ 
and relations: ($\lambda=1/\kappa$  and we use the so called ``bicrossproduct basis'')
$$
[M_k,M_l]=i\, \epsilon_{klm}M_m\,,\,\,[M_k,N_l]=i\, \epsilon_{klm}N_m\,,\,\,
[N_k,N_l]=-i\, \epsilon_{klm}M_m,$$

\begin{equation}\label{kappa-rel}           
[M_k,P_l]=i\,\epsilon_{klm}P_m\,,\,\,[M_k,P_0]=0\,,\,\,[N_k,P_0]=i\,P_k\,,\,\,[P_\mu,P_\nu]=0,
\end{equation} 
$$
[N_k,P_l]=\frac{i}{\lambda} 
\left[\frac{\delta_{kl}}{2}\left(1-\exp(-2\lambda P_0)+\lambda^2\ove{P}^2\right)-\lambda^2P_kP_l\right] $$
Now one observes that except the last relation these are exactly the relations in the Lie algebra of the Poincar\'e group, 
so it is natural to interpret $M_i$ as generators of rotations, $N_i$ as infinitesimal boosts and 
$P_\mu$ as momenta. Then  the last relation defines some ``strange'' (or deformed) action of boosts on momenta.

But due to the presence of the exponential function in the last relation they must be 
interpreted as relations in  an algebra of formal power series in  $\lambda$  and  then, 
strictly speaking, we are not allowed to assign  any  numerical value to $\lambda$. 
If we want to fix $\lambda$ we have to equip the algebra with a topology 
and interpret the exponential function as a convergent power series. 
If $\lambda$ is a number (not equal $0$) then we can remove it 
completely from relations by
defining $Y_\mu:=\lambda P_\mu$, moreover from $[N_k,P_0]=i P_k$, it follows 
that $[N_k,e^{-Y_0}]=-i\,e^{-Y_0} Y_k$,
and defining $S:=\exp(-Y_0)$  we can rewrite the relations 
(\ref{kappa-rel}) as:
$$
[M_k,M_l]=i\, \epsilon_{klm}M_m\,,\,\,[M_k,N_l]=i\, \epsilon_{klm}N_m\,,\,\,
[N_k,N_l]=-i\, \epsilon_{klm}M_m, \vspace{-4ex}$$

\begin{equation}\label{kappa-rel1}           
[M_k,Y_l]=i\,\epsilon_{klm}Y_m\,,\,\,[M_k,S]=0\,,\,\,[N_k,S]=- i\,S Y_k\,,\,\,
[Y_k ,Y_l]=0\,,\,\,[S,Y_k]=0
\end{equation} 
$$
[N_k,Y_l]=i 
\left[\frac{\delta_{kl}}{2}\left(1-S^2+ \ove{Y}^2\right)-Y_k Y_l\right] $$
The relation (\ref{dispersion}) now reads: 
\begin{equation}\frac{S^2+1-\ove{Y}^2}{2 S}=\cosh(\lambda m)=:\mu\label{def-mi}
\end{equation}

Now one can try to obtain concrete realization of these relations e.g. by operators 
on Hilbert space and ask if, and in what sense, they generate some topological algebra. 
Since momenta commute and relations among (generators of ) boosts and rotations are undeformed, 
it is natural to try to represent momenta as functions and infinitesimal 
boosts and rotations as differential  operators on the space of momenta. 
One can also ask whether this infinitesimal action  can be integrated to the action of 
the Lorentz group. And indeed this is possible as was shown  in \cite{Bruno-Ame-Glik}. 
We obtain the same result using different approach in section \ref{section-action}.
 
Having  that done, we get an action of the proper, ortochronous, Lorentz group $L_0$ 
on the subset of $R^4$:  $\tilde{V}_+:=\{(s,y)\in R_+\times R^3: s+|y|\leq1\}$ (a ``cone''). 
Orbits are hyperboloids defined by (\ref{def-mi})  
for $\mu\in [1,\infty]$. Since rotations act linearly (and only on $Y's$) and the orbit structure is
exactly as for the standard linear action of $L_0$ on 
$V_+:=\{(E,P)\in R^4 : E^2-\ove{P}^2\geq 0,\, E\geq 0\}$ 
one may suspect that this is in fact the same action
written in some strange coordinates. These  coordinates were explicitly found  in \cite{Judes}. 
By straightforward but rather tedious calculation one verifies that the mapping:
\begin{equation}\label{intertwiner}
\tilde{V}_+\ni(s,y)\mapsto 
(E,P):=\left(\frac{1-\mu s}{s\sqrt{\mu +1}}\,,\,\frac{y}{s\sqrt{\mu+1}}\right)\in V_+
\,\,,\,\,\mu=\frac{s^2+1-|y|^2}{2 s}
\end{equation}
is a homeomorphism that intertwines these two actions.

The conclusion is (putting aside some topological subtleties) that  one obtains the 
realization of relations defining $\kappa$-Poincar\'{e} algebra by  
 the crossed product of the algebra of functions on $V_+$ 
by the standard action of $L_0$. Let us denote this algebra by $\sA_+$.

After this discovery there were claims that DSR1 is nothing but the standard special relativity
\cite{Ahluwalia, Luk1} written in nonlinear coordinates on momentum space. At this point the issue of
a coproduct appears as a crucial one. The original aim of inventors of  $\kappa$-Poincar\'{e}  was
to present a deformation of the enveloping algebra of the Poincar\'{e} group with its full structure
of a Hopf algebra. Therefore the relations (\ref{kappa-rel}) are suplemented with the following 
formulae for a coproduct and an antipode:
\newcommand{\sS}{\mbox{$\cal S\,$}}
$$\Delta(P_k)=P_k\mt I + e^{-\lambda P_0}\mt P_k\,,\,\,\,\Delta(P_0)=P_0\mt I+I \mt P_0,$$
\begin{equation}\label{kappa-delta}
\Delta(M_k)=M_k\mt I+I\mt M_k\,,\,\,\,
\Delta(N_k)=N_k\mt I+e^{-\lambda P_0}\mt N_k+\lambda \epsilon_{klm} P_l\mt M_m,
\end{equation}
\begin{equation}
\label{kappa-antypod}
\sS(P_0)=-P_0\,,\,\,\,\sS(P_k)=-P_k e^{\lambda P_0}\,,\,\,\,
\sS(M_k)=-M_k\,,\,\,\,
\sS(N_k)=-e^{\lambda P_0}N_k+\lambda \epsilon_{klm}e^{\lambda P_0}P_l M_m 
\end{equation}
\kom{antypod z pracy Lukierskiego hep-th/0203065}
Or using generators (\ref{kappa-rel1}):
$$\Delta(Y_k)=Y_k\mt I + S\mt Y_k\,,\,\,\,\Delta(S)=S\mt S,$$
\begin{equation}\label{kappa-delta1}
\Delta(M_k)=M_k\mt I+I\mt M_k\,,\,\,\,
\Delta(N_k)=N_k\mt I+S\mt N_k+\epsilon_{klm} Y_l\mt M_m,
\end{equation}
\begin{equation}
\label{kappa-antypod1}
\sS(S)=S^{-1}\,,\,\,\,\sS(Y_k)=-Y_k S^{-1}\,,\,\,\,
\sS(M_k)=-M_k\,,\,\,\,
\sS(N_k)=-S^{-1} N_k+\epsilon_{klm}S^{-1} Y_l M_m 
\end{equation}

Now the standard action of $L_0$  on $V_+$ is just the restriction of the
standard action on $R^4$. Since this action is linear, one can form the semidirect product
(this is of course the Poincar\'{e} group)  in which $R^4$ and $L_0$ sit as closed subgroups. Inside
the Poincar\'{e} group we have $V_+L_0=L_0 V_+$, and any element in the  set $g\in V_+L_0$ has unique decompositions
$g=v l=\tilde{l} \tilde{v}\,\,,\, v,\tilde{v}\in V_+\,,\,l,\tilde{l}\in L_0$. Since $V_+$ is a semigroup 
in $R^4$, one can define a comultiplication on $\sA_+$ \cite{PS2}. Moreover because $R^4$ is a normal and abelian  subgroup
this comultiplication is symmetric. Restricted to functions on $V_+$ this coproduct is the one that
is defined by a semigroup structure and the group algebra of $L_0$ is contained as a {\em Hopf subalgebra}
of $(\sA_+,\Delta)$.
The conclusion is that relations (\ref{kappa-rel}) are 
compatible with a symmetric coproduct  as
it  was observed in \cite{Luk1}. But since $V_+$ is only a semigroup $\sA_+$ is not
a Hopf algebra. To have an antipode we have to add also functions on 
$V_-:=\{(E,P)\in R^4 : E^2-\ove{P}^2\geq 0,\, E\leq 0\}$, but since $V_-$ and $V_+$ generate
(as a group)  $R^4$ we would obtain the crossed product  algebra of functions on $R^4$ by the action of $L_0$. 
But the problem is that the action of $L_0$ defined on $\tilde{V}_+$ {\em cannot be extended to any larger set}.

Let's try to perform the similar analysis for the coproduct given by (\ref{kappa-delta1}). 
Firstly, one observes that $(S, Y_k)$ commute and the coproduct is as  for the algebra of
functions on the group $B:=\{(y,s)\in R^3\times R_+\}$ with the group structure:
$(y,s)(y_1,s_1)=(y+sy_1,s s_1)$.
Secondly, it is known that the $\kappa$ - deformation is formally a quantization of a
certain Lie-Poisson structure on the Poincar\'{e} group $P(4)$ which, in turn, is defined by
a  decomposition of the  Lie algebra $so(1,4)=\bb\ms so(1,3)$ (direct sum of vector spaces), 
where $\bb$ is exactly
the Lie algebra of $B$ \cite{PS3}. Now, $B$ and $L_0=SO_0(1,3)$ sit inside $SO_0(1,4)$ and we can 
ask whether there is a similar decomposition on a group level. The answer is {\em ``no, but\dots''}.
Namely, it was shown in   \cite{PS3} that one can take {\em a non connected  extension}  $A\subset SO_0(1,4)$
 of $L_0$  such that the set of decomposable elements $BA\cap AB$ is open, dense and has a full measure in 
$SO_0(1,4)$. In this situation  by the  general
construction of \cite{Vaes} one obtains a pair of   quantum groups out of this data. 
{\em Therefore the appropriate version of $\kappa$-Poincar\'{e} 
does exist in the $C^*$-algebraic setting}.
 
Let us now describe what should be ``the appropriate version'' and the results 
obtained in this paper. Since the   $\kappa$-Poincar\'{e} can be defined in any 
dimension and computations are not more complicated we will do everything in this slightly
more general setup.

Let $B:=\{(s,y)\in R_+\times R^n\}$. This is a group with a product
$(s_1,y_1) (s_2,y_2):=(s_1 s_2, s_2 y_1+y_2)$. We embed $B$ into $SO_0(1,n+1)$ 
by (one easily checks that this mapping is a group isomorphism):
\begin{equation}\label{defB} B\ni(s,y)\mapsto \left(\begin{array}{ccc} 
\frac{s^2+1+|y|^2}{2 s} & -\frac1s y^t & \frac{s^2-1+|y|^2}{2 s}\\
-y & I & -y\\
\frac{s^2-1-|y|^2}{2 s} &  \frac1s y^t & \frac{s^2+1-|y|^2}{2 s}
\end{array}\right).\end{equation}
Note that $B$ is exactly the $AN$ group that appears in the Iwasawa decomposition:
$SO_0(1,n+1)=SO(n+1)AN$.
Let  $A$ be a non-connected extension of $L_0=SO_0(1,n)$ inside $SO_0(1,n+1)$ defined by:
\begin{equation}\label{defA}
A:=\left\{\left(\begin{array}{cc} 
u & 0\\
0 & 1
\end{array}\right)
\left(\begin{array}{cc} 
I &0\\
0 & h
\end{array}\right)\,:\,\,u\in SO_0(1,n),\,h:=\left(\begin{array}{cc} 
d &0\\
0 &  d
\end{array}\right)\,,\,d=\pm 1 \right\}.
\end{equation}
To the decomposition of the  Lie algebra $so(1,n+1)=\bb\ms so(1,n)$ corresponds a local decomposition
on the level of Lie groups, $SO_0(1,n+1), B$ and $SO_0(1,n)$,  
but it is known that this decomposition is truly local  and the complement of the set of 
decomposable elements has non empty interior  \cite{Maj, PS3}. 
Yet, if we consider groups $B$ and $A$ then the set $AB\cap BA$ is open and dense in 
$SO_0(1,n+1)$ and has a full measure \cite{PS3}.

The groups $A$ and $B$ can also be defined in a more invariant  way.
Let $(V,\eta)=(R^{n+2},\eta)\,$, $\eta=diag(1,-1,\dots,-1)$ be $n+2$ dimensional Minkowski space.
Choose $v\in V$ $\eta(v,v)=1$ and let $W$ be the orthogonal complement of $v$. Choose $w\in W\,,\,\eta(w,w)=-1$
and let $U$ be the orthogonal complement of $w$ in $W$. Now $A$ and $B$ can be defined as:
$A:=\{g\in SO_0(\eta): g w=\pm w\}$, 
$B:=\{g\in SO_0(\eta): gu=u+\lambda (v-w)\,,\,g(v-w)=r (v-w)\}$ for $\lambda:SO_0(\eta)\times U\lra R$ and $r:SO_0(\eta)\lra R_+$. 

Before I summarize the main results I  want to make a comment on the variables $(s,y)$.
Variables $(s,y)$ are related to the standard DSR1 ``energy-momentum'' $(E,p)$ by 
$\lambda E=-\log s\,,\,\lambda p=y.$ As I mentioned, if we treat $\lambda$ as a fixed parameter, nothing
depends on $\lambda$, so it is better not to use it at all; ``physically'' that means we 
measure energy and momentum 
in units of $\lambda$. It is also not obvious that $(E,p)$ are really energy and 
momentum (and probably they are not).

In  section \ref{section-setup} we explicitly describe the set of decomposable elements
$AB\cap BA\subset SO_0(1,n+1)$ and corresponding projections onto $A$ and $B$.

In section \ref{section-action} we clarify what we mean by the ``action'' of $A$ on $B$, 
compute its orbits and isotropy subgroups. The orbit structure is almost exactly as in the classical situation:
``particles, antiparticles, photons and tachions'' (and one additional orbit without a 
classical counterpart). It turns out that the neutral element  of $B$ (i.e. point $(s,y)=(1,0)$) 
is a fixed point for this  ``action'' and derivative is just the standard action of  
the (restricted) Lorentz group  on $R^4$. But, as it was mentioned earlier, 
this ``action'' does not define an action of 
$SO_0(1,n)$ on any neighbourhood of the neutral element of $B$, in particular on every
``antiparticle or tachionic'' orbit there are points ``sent to infinity''. 
(It was already known to the authors of \cite{Maj} that $so(1,n)$ acts on $B$ 
by non complete vector fields). This problem
appeared also in \cite{AmeCam-Glik} in the name of ``the problem of sign of $\kappa$.'' 
On the other hand the ``action'' restricts to a true action on the  subset of 
$B$ given by $\{(s,y)\in B: s+|y|\leq 1\}$. This set is considered as the  ``momentum space'' 
in DSR1.

Section \ref{section-extension} is devoted to an extension of the ``action''. 
The manifold $\tilde{B}$ is constructed, that contains $B$ as an open and dense set
and carries a true action of group $A$. All singularities disappear. 
There is a price, however, $\tilde{B}$ is no longer a group, it is not
simply connected and in odd spacial dimension (e.g. $n=3$) non orientable. $\tilde{B}$ can
be equipped with an  $A$-invariant lorentzian metric, and locally is de Sitter space. This
is covered in section \ref{section-global}.
\section{Setup}\label{section-setup}

Let $B$  be the subgroup of a proper Lorentz group  $SO_0(1,n+1)$ 
( i.e. connected component of the identity in $SO(1,n+1)$ ) defined by (\ref{defB}) and
$A$ the  subgroup defined by (\ref{defA}).

If $n$ is odd (e.g. in dimension $1+3$) one can replace the subgroup 
$\{\left(\begin{array}{ccc} I & 0 &0 \\ 0 &d & 0\\0 & 0 & d\end{array}\right)\,,\,
d=\pm 1\}$ by the subgroup 
$\{\left(\begin{array}{ccc} 1 & 0  \\ 0 &d I\end{array}\right)\,,\,d=\pm 1\}$ , i.e. one can  parametrize $A$ by:
$$A:=\left\{\left(\begin{array}{cc} 
u & 0\\
0 & 1
\end{array}\right)
\left(\begin{array}{ccc} 
1 &0 &0\\
0&D & 0\\
0& 0 & d
\end{array}\right)\,:\,\,u\in SO_0(1,n),\,d=\pm 1\right\}\,\,,\,\,
D:=\left\{\begin{array}{cl} d\,I & n-odd\\
\left(\begin{array}{cc}I&0\\0 & d\end{array}
\right)& n-even
\end{array}\right.$$  
It is clear that $A$ is isomorphic to  the orthochronous Lorentz group.

As it was shown in \cite{PS3}  the set $\Gamma:=A B\cap B A$ is open and dense in $SO_0(1,n+1)$ 
and has a full measure.
Any element  $\gamma\in \Gamma$ can
be written in the  unique way as $\gamma=b a= a' b'$. This decomposition defines four
mappings: $a_L,a_R:\Gamma\lra A$ and $b_L,b_R: \Gamma\lra B$ by 
$b_L(\gamma)a_R(\gamma):=\gamma=:a_L(\gamma)b_R(\gamma)$. The next lemma describes
the set $\Gamma$ explicitly.
\begin{lem} {\rm 1.}  The set $\Gamma$ can be identified  with the following subset of $R^m$:
$$\{(s,y;z, U,d)\in R_+\times R^n\times R^n\times SO(n)\times\{1,-1\}:
|z|< 1\,,\,\alpha(s,y;z,U,d)\neq 0\},$$ where $\alpha$ is defined by:
$$\alpha(s,y;z,U,d):=\frac{1}{1-|z|^2} \left((1+|z|^2-2 z^t
y)+\frac{1}{2}(1-d+|z|^2(1+d))(|y|^2-1-s^2)\right)=$$
$$=\frac{1}{1-|z|^2}\left\{\begin{array}{ll}\left|\frac{z}{|z|}-|z| y\right|^2-|z|^2
    s^2 
& d=1\,,\,z\neq 0\\ 
1 & d=1\,,\,z=0\\
|z-y|^2-s^2 & d=-1\\
\end{array}\right.$$

{\rm 2.} $\G$ has four connected components, they are defined by 
$$\G_{ab}:=\{(s,y;z, U,d)\in \G : a \alpha(s,y;z, U,d)>0\,,\,b d>0\}\,,\,a,b=+,-.$$
\end{lem}

\begin{re} The function $\alpha$ has a geometric meaning, it is related to various modular functions
defined on the set $AB\cap BA$. In notation of \cite{PS2} $\alpha^n=Q$.
\end{re}
\noindent
{\em Proof:} 1. We parameterize $SO_0(1,n)$ by: 
$$\{z\in R^n: |z|<1\}\times SO(n)\ni (z,U)\mapsto \left(\begin{array}{cc}
\frac{1+|z|^2}{1-|z|^2} & \frac{2}{1-|z|^2}z^tU\\ 
\frac{2 }{1-|z|^2} z & (I+\frac{2}{1-|z|^2}z z^t)U\end{array}\right)\in
SO_0(1,n)$$
This is the standard  $lorentz=boost\times rotation$ but because of computational reasons
we use $z$ instead of velocity $v$. They are related by:
$$z=\frac{v}{1+\sqrt{1-|v|^2}}\,\,,\,\,\,v=\frac{2 z}{1+|z|^2}$$
In this way we obtain a parameterization of $BA$:
$$(s,y;z,U,d)\mapsto \left(\begin{array}{ccc} 
\frac{s^2+1+|y|^2}{2 s} & -\frac1s y^t & \frac{s^2-1+|y|^2}{2 s}\\
-y & I & -y\\
\frac{s^2-1-|y|^2}{2 s} &  \frac1s y^t & \frac{s^2+1-|y|^2}{2 s}
\end{array}\right) \left(\begin{array}{ccc}
\frac{1+|z|^2}{1-|z|^2} & \frac{2}{1-|z|^2}z^tU & 0\\ 
\frac{2 }{1-|z|^2} z & (I+\frac{2}{1-|z|^2}z z^t)U & 0\\
0 & 0 & 1\end{array}\right) \left(\begin{array}{ccc} 
I& 0 &0\\
0 & D & 0\\
0& 0 & d
\end{array}\right)$$

Now it is clear that $(s,y;z,U,d)\in \Gamma$ if and only if there exist 
$(\tilde{s},\tilde{y};\tilde{z},\tilde{U},\tilde{d})$ such that the following
equation is satisfied:
\begin{eqnarray} \label{basic-eq}\nonumber
\left(\begin{array}{ccc} 
\frac{s^2+1+|y|^2}{2 s} & -\frac1s y^t & \frac{s^2-1+|y|^2}{2 s}\\
-y & I & -y\\
\frac{s^2-1-|y|^2}{2 s} &  \frac1s y^t & \frac{s^2+1-|y|^2}{2 s}
\end{array}\right) 
\left(\begin{array}{ccc}
\frac{1+|z|^2}{1-|z|^2} & \frac{2}{1-|z|^2}z^tU & 0\\ 
\frac{2 }{1-|z|^2} z & (I+\frac{2}{1-|z|^2}z z^t)U & 0\\
0 & 0 & 1\end{array}\right) 
\left(\begin{array}{ccc} 
1& 0 &0\\
0 & D & 0\\
0& 0 & d
\end{array}\right)=\\
=\left(\begin{array}{ccc}
\frac{1+|\tilde{z}|^2}{1-|\tilde{z}|^2} & 
\frac{2}{1-|\tilde{z}|^2}\tilde{z}^t\tilde{U} & 0\\ 
\frac{2 }{1-|\tilde{z}|^2} \tilde{z} & 
(I+\frac{2}{1-|\tilde{z}|^2}\tilde{z} \tilde{z}^t)\tilde{U} & 0\\
0 & 0 & 1\end{array}\right) 
\left(\begin{array}{ccc} 
1& 0 &0\\
0 & \tilde{D} & 0\\
0& 0 & \tilde{d}
\end{array}\right)\left
(\begin{array}{ccc} 
\frac{\tilde{s}^2+1+|\tilde{y}|^2}{2 \tilde{s}} & -\frac{1}{\tilde{s}} \tilde{y}^t & 
\frac{\tilde{s}^2-1+|\tilde{y}|^2}{2 \tilde{s}}\\
-\tilde{y} & I & -\tilde{y}\\
\frac{\tilde{s}^2-1-|\tilde{y}|^2}{2 \tilde{s}} &  \frac{1}{\tilde{s}} \tilde{y}^t & 
\frac{\tilde{s}^2+1-|\tilde{y}|^2}{2 \tilde{s}}
\end{array}\right),
\end{eqnarray}
 
This equation can be solved (see Appendix) 
 and it turns out that the condition $\alpha(s,y;z,U,d)\neq0$ 
is necessary and sufficient for existence of solution.\vs\\
2.  It is clear that $\G$ has at least four connected components. So
 we have to show that each $\G_{ab}$ is connected.
\begin{itemize}
\item  $\G_{++}=\{(s,y;z,U,1): \alpha>0\}$. Of course we can put  $U=I$. For fixed
  $z\neq 0$ the set $\{(s,y):\alpha(s,y;z,I,1)>0\}$ is the exterior of the cone
$|y-\frac{z}{|z|^2}|=s$, so it is connected (for $n>1$). Moreover $(1,0)$ is in this
set. Thus $\G_{++}$ is connected.
\item $\G_{-+}=\{(s,y;z,U,1):\alpha<0\}$. Again we can assume that $U=I$. Now, for
fixed $z$ we have interior of the cone, so connected set. For any two $z_1,z_2$ these
cones have non empty intersection.
\item In the same way one proves that the  sets $\G_{+-}=\{(s,y;z,U,-1):|y-z|>s\}$
  and $\G_{--}=\{(s,y;z,U,-1): |y-z|<s\}$ are connected. 
\end{itemize}\dowl\vs\\
Solution of the equation (\ref{basic-eq}) gives us the explicit form of the mappings
$a_L,\,a_R,\,b_R,\,b_L$.
\begin{equation}\label{bl-ar} b_L(s,y;z,U,d)=(s,y)\,,\ a_R(s,y;z,U,d)=(z,U,d)\end{equation}
and $a_L(s,y;z,U,d)=:(\tilde{z},\tilde{U},\tilde{d})=(\tilde{z},\tilde{U},sgn(\alpha))$ where
$(\tilde{z}, \tilde{U})$ are given by:\\
For  $(s,y;z,U,d)\in\G_{++}$ (i.e. $\alpha>0\,,d=1$):
\begin{eqnarray}\label{al++}
\tilde{z}=\frac{s |z|^2}{|z-|z|^2y|^2}(z-|z|^2 y)\,\,,\,\,
\tilde{U}=\left[I+\frac{2 |z|^2}{|z-|z|^2 y|^2}(-|y|^2 z z^t+ z y^t+(-1+2 z^t y) y
  z^t-|z|^2 y y^t)\right] U;\end{eqnarray}
For  $(s,y;z,U,d)\in\G_{-+}$ (i.e. $\alpha<0\,,d=1$):
\begin{equation}\label{al-+} 
\tilde{z}=\frac{1}{s |z|^2}(z-|z|^2 y)\,,\,\,
\tilde{U}=\left[I-\frac{2}{ |z|^2}z z^t\right] U\tilde{D};
\end{equation}
For  $(s,y;z,U,d)\in\G_{--}$ (i.e. $\alpha<0\,,d=-1$):
\begin{equation}\label{al--}
\tilde{z}=\frac{1}{s} (z-y)\,,\,\,\tilde{U}=U;
\end{equation}
For  $(s,y;z,U,d)\in\G_{+-}$: (i.e. $\alpha>0\,,d=-1$):
\begin{equation}\label{al+-}
\tilde{z}=\frac{s}{|z-y|^2} (z-y)\,,\,\,
\tilde{U}=\left[I-\frac{2}{|z-y|^2}(z-y) (z-y)^t\right]U D;
\end{equation}
And the right projection on $B$: $b_R(s,y;z,U,d)=:(\tilde{s},\tilde{y})$ 
where  $(\tilde{s},\tilde{y})$ are given by: 
\\
For  $(s,y;z,U,d)\in\G_{++}\cup\G_{-+}$ (i.e. $d=1$):
\begin{eqnarray}\label{br+}
\tilde{s}& = & \frac{s}{|\alpha|}=\frac{s (1-|z|^2)|z|^2}{||z-|z|^2y|^2-|z|^4s^2|},\\
\nonumber \tilde{y}& = & \frac{|z|^2 U^{-1}}{|z-|z|^2y|^2-|z|^4s^2}
\left((s^2-1-|y|^2+2 z^t y) z +(1-|z|^2) y\right)=
\frac{U^{-1}}{\alpha}\left((s^2-|y|^2-\alpha) z+y\right);
\end{eqnarray}
For $(s,y;z,U,d)\in\G_{+-}\cup\G_{--}$: (i.e. $d=-1$)
\begin{eqnarray}\label{br-}
\tilde{s}& = & \frac{s}{|\alpha|}=\frac{s (1-|z|^2)}{||z-y|^2-s^2|},\\
\nonumber \tilde{y}& = &\frac{D U^{-1}}{|z-y|^2-s^2}\left((s^2-1-|y|^2+2 z^t y) z +(1-|z|^2)
  y\right)=\frac{D U^{-1}}{\alpha}\left((-1-\alpha)z +y\right).
\end{eqnarray}

In what follows we will write  $z$ for  $(z,I,1)\in A$, $U$ for   
$(0,U,1)\in A$ and   $J$ for $(0,I,-1)\in A$. We define  $A_+:=\{(z,U,d)\in A: d=1\}=SO_0(1,n)$, 
$A_-:=\{(z,U,d)\in A: d=-1\}=A_+ J$,  we will also omit $U$ in arguments of $\alpha$;
\section{``Action'' of $A$ on $B$.}\label{section-action}

In this section we decribe in what sense $A$ ``acts'' on $B$ and find orbits of
this ``action''. The groups $A,B$ and $G:=SO_0(1,n)$ satisfy the assumptions of 
the following lemma:
\begin{lem}\label{action}
\notka{action}
Let $G$ be a Lie group and  $A,B\subset G$ closed subgroups. Assume  that the set 
$\Gamma:=A B \cap B A$ is open and dense in $G$, and for every $a\in A$ the set 
$D_a:=\{b\in B\,:\,b a \in \Gamma\}$ is open and dense in
$B$, and  for every $b\in B$ the set $D_b:=\{a\in A\,:\,b a \in \Gamma\}$ is 
open and dense in $A$.\\ 
For $a\in A$ let us define a mapping: 
$$\phi_a: D(\phi_a):=D_a\ni b\mapsto \phi_a(b):=b_R(b a)\in B.\vsm$$
Then:\vsm
\begin{enumerate}
\item For every $a\in A$ the mapping $\phi_a$ is a diffeomorphism from 
$D_a$ onto $D_{a^{-1}}$,  and $\phi_{a^{-1}}\phi_a(b)=b$ for $b\in
D_a$;\vsm
\item For all $a,a'\in A$ the set $D(\phi_{a}\phi_{a'})$ is open and dense in 
$D_{a' a}$, and $\phi_{a}\phi_{a'}(b)=\phi_{a' a}(b)$ for $b\in
D(\phi_{a}\phi_{a'})$.
\end{enumerate}
\end{lem}
{\em Proof:} 1) It is clear that $\phi_a$ is smooth. For $b\in D_a$ there exists
$a'\in A$ such that  $ba=a' \phi_a(b)$. So  $(a')^{-1} b=
\phi_a(b) a^{-1}$ i.e. $\phi_a(b)\in D_{a^{-1}}$ and $\phi_{a^{-1}}\phi_a(b)=b$.\\
2) The condition $b\in D(\phi_a\phi_{a'})$ and $b''=\phi_a\phi_{a'}(b)$ is equivalent to
$$\exists \, a'',a'''\in A,\, b',b''\in B\,: ba'=a''b'\,,\,b'a=a'''b''$$
so $ba'a=a''b'a=a''a'''b''$ and $b\in D_{a'a}$ and $\phi_{a'a}(b)=b''$. 
On the other hand $b\in D(\phi_a\phi_{a'})$ means 
$b\in D_{a'}\cap \phi_{a'}^{-1}(D_a\cap D_{a'^{-1}})$ and this set is open and dense
in $B$ so in $D_{a'a}$.\\
\dowl\vs\\
In our situation for $a:=(z,U,d)$ we have  $D_a:=\{(s,y): \alpha(s,y;z,d)\neq
0\}$, and we have an ``action'' (in the sense of the previous lemma) 
$\phi_a : D_a\lra B\,,\, \,\phi_a(s,y):=b_R(s,y;z,U,d)$.\\
Later on we will need the derivatives of mappings $\phi_a$. 

\begin{lem} \label{der}\notka{der}
i) In the expression below $\alpha=\alpha(s,y;z,1)$ and $(s,y)\in D_z$
$$\phi_z'(s,y)=\frac{1}{\alpha^2(1-|z|^2)}\times $$
$$\times \left(\begin{array}{cc}sgn(\alpha)\left(\left|\frac{z}{|z|}-|z|
        y\right|^2+|z|^2 s^2\right) & 2 s\,sgn(\alpha) (z^t-|z|^2 y^t)\\
2 s[(1-2 z^t y) z +|z|^2 y] & \alpha(1-|z|^2)I+2 (s^2-|y|^2) z z^t+ 2 y z^t-2 (1-2
z^t y) z y^t-2 |z|^2 y y^t\end{array}\right),$$
$$\det(\phi_z'(s,y))=sgn(\alpha)\alpha^{-n-1};$$
ii) In the expression below $\alpha=\alpha(s,y;0,-1)$ and $(s,y)\in D_J$
$$\phi_J'(s,y)=\frac{1}{\alpha^2}\left(\begin{array}{cc}sgn(\alpha)(|y|^2+s^2) & -2 s
    sgn(\alpha) y^t\\ 2 s D y & \alpha D-2 Dy y^t\end{array}\right),$$
$$\det(\phi_J'(s,y))=sgn(\alpha)\alpha^{-n-1};$$
iii) $\phi_U'(s,y)=\left(\begin{array}{cc}1 & 0 \\ 0 & U^t\end{array}\right);$
\end{lem}
{\em Proof:} Computations.\dowl\\
Notice that $\phi_z'(1,0)$ is a matrix of an ordinary boost and 
$\phi_J'(1,0)=\left(\begin{array}{cc} -1 & 0\\0&-D\end{array}\right)$.\vs\\

Now we check that we really have obtained the realization of the commutation relations (\ref{kappa-rel1}).
Let $\{\Lambda_{\alpha\beta}\,,\,\alpha<\beta\,,\,\alpha,\beta=0,\dots, n\}$ be the standard basis of the Lie algebra 
$so(1,n)$. Define the vector fields 
$$\tilde{N}_k(s,y):=\left.\frac{d}{d t}\right|_{t=0}b_R(s,y;\exp(t\Lambda_{0k}),I,1)\,\,{\rm and}\,\,\,
\tilde{M}_{kl}(s,y):=\left.\frac{d}{d t}\right|_{t=0}b_R(s,y;0,\exp(t\Lambda_{kl}),1)\,,\,\,k,l=1,\dots,n.$$
By the definition they satisfy commutation relations for $so(1,n)$. The explicit form of these vector fields is
$$\tilde{N}_k(s,y)=s y_k \partial_s+\sum_{j=1}^{n}(y_k y_j+\frac12(s^2-|y|^2-1)\delta_{kj})\partial_{y_j} \,\,{\rm and}\,\,\,
\tilde{M}_{kl}(s,y)=-(y_k\partial_{y_l}-y_l\partial_{y_k}).$$
Finally, let $N_k:=-i\tilde{N}_k$, $M_{kl}:=-i \tilde{M}_{kl}$, $S$ be a multiplication by $s$ and $Y_k$ multiplication by 
$y_k$ (all operators act on $C^{\infty}(B)$). Straightforward computations show that these operators satisfy
(appropriately generalized to higher dimension) relations (\ref{kappa-rel1}).

As for a true  action, the relation 
$\{(b_1,b_2)\in B\times B : \exists\, a\in A\,: b_1=\phi_a(b_2)\}$ is an equivalence
relation and $B$ is a disjoint union of orbits.
Let us define the following subsets of $B$ (see the picture): 
\notka{defregions}
\begin{eqnarray}\label{defregions}
H_\mu=\{(s,y)\in B\setminus\{(1,0)\}: \frac{s^2+1-|y|^2}{2 s}=\mu\}\,,\,\mu\in
R;\nonumber\\
H_\mu^+:=\{(s,y)\in H_\mu: s<1\}\,\,,\,\,\,
H_\mu^-:=\{(s,y)\in H_\mu: s>1\}\nonumber\\
BI:=\{(s,y)\in B: s-|y|<-1\}\,\,\,,\,\,\,\,
BII:=\{(s,y)\in B: -1 <s-|y|< 1\,,\,s+|y|>1\}, \nonumber\\
BIII:=\{(s,y)\in B: s-|y|> 1\}\,\,,\,\,\,\,BIV:=\{(s,y)\in B: s+|y|<1\}.
\end{eqnarray}
\begin{lem}\label{rozklad}
\notka{rozklad}
\begin{itemize}
\item $BI=\bigcup_{\mu\in ]-\infty,-1[}\, H_\mu$;
\item $BII=\bigcup_{\mu\in ]-1,1[}\, H_\mu$;
\item $BIII=\bigcup_{\mu\in ]1,\infty[} \,H_\mu^-$;
\item $BIV=\bigcup_{\mu\in ]1,\infty[} \,H_\mu^+ $;
\item $B=BI\cup BII\cup BIII\cup BIV\cup H_1\cup H_{-1}\cup\{(1,0)\}$ (disjoint union).
\end{itemize}
\end{lem}
\kom{Ten lemat zostal sprawdzony.}
{\em Proof:} Direct computations.\dowl\\
Let $\mu$ be the function:
\notka{defmi}
\begin{equation}\label{defmi}
B\ni (s,y)\mapsto \mu(s,y):=\frac{s^2+1-|y|^2}{2 s}\in R.
\end{equation}
Straightforward computations show that:
\notka{fi-mi}
\begin{equation}\label{fi-mi}
\mu(\phi_a(s,y))=d\, sgn(\alpha(s,y;z,d))\, \mu(s,y)\,,\,\,a:=(z,U,d)\,,\,(s,y)\in D_a
\end{equation}
\begin{center}
\begin{minipage}{0.9\textwidth}
\newgray{grey}{1}
\newgray{grey1}{0.5}
\newgray{grey2}{0.5}
\psset{xunit=0.22\textwidth}
\psset{yunit=0.22\textwidth}
\begin{pspicture}(0,0)(4.2,4.2)
\psaxes[linewidth=1.5pt,labels=none]{->}(0,0)(4,4)
\uput[l](0,4){\Large $s$}
\uput[d](4,0){\Large $|y|$}
\pspolygon[linestyle=none,fillstyle=solid,fillcolor=grey](0,1)(1,0)(4,3)(4,4)(3,4)
\pspolygon[linestyle=none,fillstyle=vlines,hatchangle=90,hatchcolor=grey1,hatchsep=12pt](0,1)(3,4)(0,4)
\pspolygon[linestyle=none,fillstyle=vlines,hatchangle=90,hatchcolor=grey1,hatchsep=12pt](0,1)(1,0)(0,0)
\pspolygon[linestyle=none,fillstyle=hlines,hatchangle=90,hatchcolor=grey1,hatchsep=12pt](1,0)(4,3)(4,0)
\psline(0,1)(3,4)
\psline(0,1)(1,0)
\psline(1,0)(4,3)
\psline[linestyle=dotted](0,0)(4,4)
 \psplot[linewidth=2.pt]{1}{4}{x 2 exp 1 sub sqrt} 
 \psplot{0}{2}{x 2 exp 1.25 add sqrt 1.5 add} 
 \psplot{0}{1}{1.5 x 2 exp 1.25 add sqrt sub} 
 \psplot{1}{4}{x 2 exp 3 add sqrt -2 add} 
 \psplot{1}{3}{x 2 exp 0.75 sub sqrt -0.5 add} 
 \psplot{0.6}{1}{0.8 x 2 exp 0.36 sub sqrt  sub} 
 \psplot{0.6}{3}{0.8 x 2 exp 0.36 sub sqrt  add} 
\rput(0.3,3.2){\Large $\mu>1$}
\rput(0.2,0.55){\Large $\mu>1$}
\rput(0.2,0.22){\Large $H_\mu^+$}
\rput(1,1){\large $1>\mu>0$}
\rput(2.5,0.5){\Large $\mu<-1$}
\rput(3.1,2.5){\large $0>\mu>-1$}
\rput(3.5,3.2){\Large $H_0$}
\rput(1.3,3.05){\Large $H_\mu^-$}
\end{pspicture}
\end{minipage}
\end{center}
\vspace*{5ex}

Now we can describe the orbits of our ``action''. 
\begin{prop}\label{orbits}
\notka{orbits}
\begin{enumerate}
\item There are the following orbits  of  the  ``action'' of $A$ on $B$:
\begin{enumerate}
\item  1-point orbit $\{(1,0)\}$, 
\item $H_\mu\cup H_{-\mu}$ for $\mu\in R$;
\end{enumerate}
\item There are the following orbits  of the ``action'' of $A_+(=SO_0(1,n))$ on $B$:
\begin{enumerate}  
\item 1-point orbit $\{(1,0)\}$, 
\item $H_\mu^+$ for $\mu\geq 1$, 
\item $H_\mu^-\cup H_{-\mu}$ for $\mu\geq 1$,
\item $H_\mu\cup H_{-\mu}$ for $0\leq \mu <1$, 
\end{enumerate}
\item The ``action'' of $A_+$ restricted to the set $BIV\cup H_1^+$ 
(manifold with a boundary) defines an action of $A_+$. 
\end{enumerate}
\end{prop}
{\em Proof:} We start by proving (3). Let $a=(z,U,1)\in A_+$. 
For  $s+|y|\leq 1$ we have $s^2\leq (1-|y|)^2$ and 
$$(1-|z|^2) \alpha(s,y;z,1)=
1-2z^t y+|z|^2|y|^2-|z|^2s^2\geq 1-2z^t y+|z|^2|y|^2-|z|^2(1-2|y|+|y|^2)= $$
$$=1-2z^t y-|z|^2+2 |z|^2|y|\geq
1-2|z||y|-|z|^2+2|z|^2|y|=(1-|z|)(1+|z|-2|z||y|)\geq$$
$$\geq(1-|z|)(1+|z|-2|z|)=(1-|z|)^2>0$$
i.e. this set is contained in $D_a$.\\
By the  lemma \ref{action} the image of this set is also  contained in the domain of every
$\phi_a$. On the other hand,  if $s+|y|>1, y\neq 0$, 
it is easy to check that the point $(s,y)$ is
not in the domain of $\phi_z$ for $z:=\frac{1}{|y|( |y|+s)}\, y$. For
$(s,0)\,,\,s>1$ the point $(s,0)$ is not in the domain of 
$\phi_z$ for $z:=\frac{1}{s|x|}\, x\,,\,x\in R^n\setminus\{0\}$. That means the
``action'' of $A_+$ restricted to this set is a true action. It is clear that the map: 
$\,(BIV\cup H_1^+)\times A_+\ni(s,y;z,U,1)\mapsto b_R(s,y;z,U,1)$ is smooth. 
Point (3) is proven.
\begin{re}
This is the set considered in DSR1 as the momentum space with the action of the (proper)
Lorentz group.
\end{re}
Now we will prove points (1) and (2). First observe that by (\ref{fi-mi}):
$$\phi_a((H_\mu\cup H_{-\mu})\cap D_a)\subset H_\mu\cup H_{-\mu}$$
i.e. $H_\mu\cup H_{-\mu}$ are invariant.\\
{\em (1a) and (2a): }
Using the formulae defining ``action'' one immediately checks that
$(1,0)\in D_a$ for any $a\in A$, and $(1,0)$ is a fixed point of $\phi_a$.\\
Next lemma describes the ``action'' of boosts.
\begin{lem}\label{restframe}
\notka{restframe}
{\em a)} For every $(s,y)\in BI\cup BIII\cup BIV$ there exists exactly one
pair $(\tilde{s},z)$ such that $(s,y)=\phi_z(\tilde{s},0)$. It is 
given by ( $\mu:=\mu(s,y)=\frac{s^2+1-|y|^2}{2 s}$ )
\begin{itemize}
\item For $(s,y)\in BI$: 
$\tilde{s}:=-\mu+\sqrt{\mu^2-1}>1\,\,,\,\,\,z:=-\frac{1}{1+\tilde{s} s}\,y$ 
( $z^t y<0$ );
\item  For $(s,y)\in BIII$: $\tilde{s}:=\mu+\sqrt{\mu^2-1}>1 \,\,,\,\,\,z:=\frac{1}{\tilde{s}
  s-1}\,y$ (  $z^t y>0$ );
\item For $(s,y)\in BIV$: $\tilde{s}:=\mu-\sqrt{\mu^2-1}<1\,\,,\,\,\,
z:=\frac{1}{\tilde{s} s-1}\,y$ ($z^t y<0$ ).
\end{itemize}
{\em b)} For every $(s,y)\in BII\cup H_{-1}\cup H_1^-$ there exists exactly
one pair $(\tilde{y},z)$ such that $(s,y)=\phi_z(1,\tilde{y})$. It is 
given by 
\begin{itemize}
\item for  $\,-1\leq \mu\leq 0$: $\tilde{y}:=\sqrt{2 (1-\mu)}\frac{y}{|y|}\,,\,
z:=\frac{s-1}{|y|+s \sqrt{2 (1-\mu)}}\frac{y}{|y|}$,
\item for $\,0<\mu\leq 1$: $\tilde{y}:=-\sqrt{2 (1+\mu)}\frac{y}{|y|}\,,\,
z:=-\frac{s+1}{|y|+s \sqrt{2 (1+\mu)}}\frac{y}{|y|}$.
\end{itemize}
Note that in both cases $-1\leq \mu(1,\tilde{y})\leq 0$.\\
{\em c)} For every $(s,y)\in H_1^+$  there exists exactly one $(\tilde{y},z)$
such that $\,(s,y)=\phi_z(1/2,\tilde{y})$. It is given by
$\tilde{y}=\frac{y}{2 |y|}\,,\,z:=(2 s -1)\,\frac{y}{|y|}.$
\end{lem}
{\em Proof:} Computation.\\\dowl\\
{\em (2b)}: It follows from point  (3) of the proposition together with (c) and third statement of (a)  
of the lemma \ref{restframe}. \\
{\em (2c)}:
For $\mu>1$ this follows from the first and the second statement of the point (a) of the lemma. 
For $\mu=1$ use point (b) of the lemma.\\
{\em (2d)}: It follows from point (b) of the lemma.\\
{\em (1b)}: It is enough to prove that for $\mu\geq 1$ $H_\mu^+$ and $H_\mu^-$ 
are contained the same orbit. But straightforward computation shows that
$\phi_J(H_\mu^+)=H_\mu^- \cup H_{-\mu}$ for $\mu\geq 1$.
\\\dowl\vs

Now  we compute the isotropy groups. Let $G_x$ be the isotropy group of $x$ in $A$ and
$G_x^+:=G_x\cap A_+$ be the isotropy group in $A_+$. Of course the isomorphism class of $G_x$ 
depends only on $A$-orbit of $x$. From the previous proposition  and the fact that for 
$a\in A_+$ $J a J\in A_+$ it follows that  also isomorphism class of $G_x^+$ depends only on 
$A$-orbit of $x$, i.e. only on $|\mu|$. In the following proposition $G_\mu$ and $G_\mu^+$ denote 
corresponding abstract groups and, for $a\in A$ and $x\in R^{n+1}$, $a x$ denotes the standard action
of $A$ on Minkowski space $(R^{n+1},\eta)$.
\begin{prop}\label{isotropy}
There are the following isotropy groups\\
a) for $|\mu|>1$:  $G_\mu\simeq G_\mu^+\simeq\{a\in A_+ :  a x=x\,,\,\eta(x,x)>0\}$; \\
b) for $|\mu|=1$:   $G_\mu\simeq \{a\in A_+ : a x= x\,,\,\eta(x,x)=0\}\cup \{a\in A_- : a x= -x\,,\,\eta(x,x)=0\}$;\\
c) for $0<|\mu|<1$:  $G_\mu\simeq \{a\in A_+ : a x= x\,,\,\eta(x,x)<0\}\cup \{a\in A_- : a x= -x\,,\,\eta(x,x)<0\}$;\\
d) for $\mu=0$:  $G_\mu\simeq \{a\in A: a x=\pm x\,,\,\eta(x,x)<0\}$.
\end{prop}

\noindent
{\em Proof:} a) It is enough to compute the isotropy of $(s,0), s<1$, this is straightforward. 
Here $x:=\left( \begin{array}{c}s\\0\end{array}\right)$;\\
b) It is enough to compute the isotropy of $(\frac{1}{2}, y), |y|=\frac{1}{2}$. 
Here $x:=\left( \begin{array}{c}|y|\\y\end{array}\right)$;\\
c) Here one can assume that $s=1$, $0< |y|^2<2$, and one gets $x:=\left( \begin{array}{c} 1-\mu \\y\end{array}\right)$;\\
d) Here one can assume that $s=1$, $|y|^2=2$, and  one gets $x:=\left( \begin{array}{c} 1 \\y\end{array}\right)$.\\
\dowl

Using the interpretation of coordinates $(s,y)$ in the region $BIV$ adopted in DSR1 
(i.e. $E=-\log s\,,\,p=y$) one may  call orbits $H_\mu^+\,,\,\mu\geq 1$ ``particle'' orbits.
($\mu=1$ corresponds to photons). Lemmas \ref{restframe}, \ref{isotropy} suggest that we should call
$H_\mu^-\cup H_{-\mu}\,,\,\mu>1$ ``antiparticle'' orbit and $H_\mu\cup H_{-\mu}\,,\,0\leq\mu<1$
``tachion'' orbit. But for these orbits the action of some boosts is not  defined and they
are disconnected. In the next section we show the way to remove this inconvenience and
we will get the orbits (almost) exactly as in the classical situation.
\section{Extension of the ``action''}\label{section-extension}

The manifold $B$ will  be extended to a new manifold $\tilde{B}$.
$B$ is open and dense in $\tilde{B}$ and $A$ acts smoothly on $\tilde{B}$.

On the set $R\times S^{n-1}$  let us define the
equivalence relation: $(\mu, x)\sim (-\mu,-x)$ and let $W:=(R\times S^{n-1})/\sim$.
Define $\tilde{B}:=B\cup W$, and for $(z,U,d)\in A$ the mapping 
$\phi(z,U,d):\tilde{B}\lra\tilde{B}$ by the formulae:\vs\\
$\phi(z,U,1):$\nopagebreak
\renewcommand{\arraystretch}{1.2}
\notka{ext+1}
\begin{eqnarray}\label{ext+1}\nonumber\nopagebreak
B\ni(s,y)\mapsto\left\{
\begin{array}{lcl}
\left(\frac{s}{|\alpha|}\,,\,\frac{U^{-1}}{\alpha}((s^2-|y|^2-\alpha) z+y)\right) \in B
& {\rm\,for\,} & \alpha:=\alpha(s,y;z,1) \neq 0 \\  
 \ [ (\mu(s,y)\,,\,\frac{U^{-1}}{s}((s^2-|y|^2)z+y))]   \in W & {\rm \, for \,} & \alpha=0
\end{array}\right.
\\\\
\nonumber
W\ni[(\mu,x)]\mapsto \left\{
\begin{array}{lcl}
\left(\frac{1-|z|^2}{2 |\mu |z|^2+z^t x|}\,,\,
\frac{-U^{-1}}{\mu |z|^2+z^t x}(\frac{1-|z|^2}{2} x+(\mu+z^t x) z)\right)\in B & {\rm\,for\,} &
\mu |z|^2+z^t x\neq 0 \\
 \ [(\mu\,,\,U^{-1}(x+2 \mu z))]\in W & {\rm\,for\,} & \mu |z|^2+z^t x = 0
\end{array}\right.
\end{eqnarray}
$\phi(z,U,-1):$
\notka{ext-1}
\begin{eqnarray}\label{ext-1}\nonumber
B\ni(s,y)\mapsto\left\{
\begin{array}{lcl}
\left(\frac{s}{|\alpha|}\,,\,\frac{D U^{-1}}{\alpha}(y- (1+\alpha) z )\right) \in B
& {\rm\,for\,} & \alpha:=\alpha(s,y;z,-1) \neq 0 \\  
 \ [ (- \mu(s,y)\,,\,\frac{D U^{-1}}{s}(y-z))]   \in W & {\rm \, for \,} & \alpha=0
\end{array}\right.
\\\\
\nonumber
W\ni[(\mu,x)]\mapsto \left\{
\begin{array}{lcl}
\left(\frac{1-|z|^2}{2 |\mu+z^t x|}\,,\,
\frac{-D U^{-1}}{\mu +z^t x}(\frac{1-|z|^2}{2} x+(\mu+z^t x) z)\right)\in B & {\rm\,for\,} &
\mu +z^t x\neq 0 \\
 \ [(-\mu\,,\,D U^{-1}x)]\in W & {\rm\,for\,} & \mu +z^t x = 0
\end{array}\right.
\end{eqnarray}
By straightforward but rather tedious computations one proves:
\begin{prop}\label{ext-action}
\notka{ext-action}
The mapping $\Phi:\tilde{B}\times A\ni(\tilde{b},a)\mapsto \phi(a)(\tilde{b})\in \tilde{B}$ is
a right action of $A$ on $\tilde{B}$.
\end{prop}\dow

Now we are going to equip $\tilde{B}$ with a structure of a smooth manifold and
to prove that the action is smooth.\\
Let $O_0:=B,\varphi_0:=id$ and for $k=1,2,\dots,n$ let
$O_k:=\{(\sigma,\xi)\in R\times R^n: \xi_k>0\,,\,\frac12< |\xi| <2\}$, and 
$\varphi_k:O_k\lra\tilde{B}$ be defined as:
\notka{def-fik}
\begin{equation}\label{def-fik}
\varphi_k(\sigma,\xi):=\left\{\begin{array}{lcr} 
[(\sigma,\xi)]\in W  & {\rm for} &|\xi|=1\\
\left(\frac{\sigma(1-|\xi|)+\sqrt{\sigma^2(1-|\xi|)^2+2|\xi|-1}}{|1-|\xi||} 
\,,\,\frac{\xi}{1-|\xi|}\right)\in B
& {\rm for} &|\xi|\neq 1\\
\end{array}\right.
\end{equation}
\begin{lem}\label{atlas}
\notka{atlas} 
For $k,l=0,\dots,n$:\\
1) $\varphi_k$ is a bijection $O_k\lra \varphi_k(O_k)$;\\
2) $\varphi_0(O_0)\cup\dots\cup\varphi_n(O_n)=\tilde{B}$;\\
3) $\varphi_k^{-1}\varphi_l$ are smooth.
\end{lem}
{\em Proof:} 1) Let $\tilde{O}_k:=\{[(\mu,x)]\in W : x_k\neq 0\}\cup \{(s,y)\in B :
(y_k>0, |y|>1) {\rm \,or\, } (y_k<0 , |y|>2\}$ and define the mapping 
$\tilde{\varphi}_k: \tilde{O}_k\lra R\times R^n$:
$$[(\mu,x)]\mapsto (sgn(x_k) \mu\,,\,sgn(x_k) x);$$
$$(s,y)\mapsto\left\{\begin{array}{lcr} 
(\mu(s,y)\,,\, \frac{y}{1+|y|}) & {\rm for} &|y|>1\,,\,y_k>0\\
(-\mu(s,y)\,,\, \frac{y}{1-|y|}) & {\rm for} &|y|>2\,,\,y_k<0
\end{array}\right.$$
One easily checks that $\tilde{O}_k=\varphi_k(O_k)$ and $\tilde{\varphi}_k=\varphi_k^{-1}$;\\
2) This is clear from the definition of $\varphi_k$;\\
3) Let  $l,k>0$. By direct calculation one obtains: 
$$\varphi_l^{-1}(\varphi_k(O_k)\cap\varphi_l(O_l))=$$\vsm\vsm\vsm\vsm
$$=\{(\sigma,\xi): \xi_l>0,\xi_k>0, \frac12<|\xi|<2\}\cup
\{(\sigma,\xi): \xi_l>0,\xi_k<0, \frac23<|\xi|<2\}=:D_{++}\cup D_{+-}.$$
So the domain of $\varphi_k^{-1}\varphi_l$ is open and 
$$\varphi_k^{-1}\varphi_l(\sigma,\xi)=\left\{\begin{array}{lcr} 
(\sigma\,,\, \xi) & {\rm for} &(\sigma,\xi)\in D_{++}\\
(-\sigma\,,\, \frac{\xi}{1-2 |\xi|}) & {\rm for} &(\sigma,\xi)\in D_{+-}
\end{array}\right.$$
For $k=0$ we have $\varphi_0(U_0)\cap \varphi_l(U_l)=B\cap \varphi_l(U_l)$ and
$\varphi_0^{-1}\varphi_l=\varphi_l$ on the domain $\{(\sigma,\xi)\in U_l : |\xi|\neq 1\}$, and
$\varphi_l^{-1}\varphi_0=\varphi_l^{-1}$ on the domain $\tilde{U}_l\cap B$ and 
these mappings are smooth.
\\\dowl\\
In this way we have defined the structure of a $C^\infty$-manifold on $\tilde{B}$ such that
$B$ is open and dense  in $\tilde{B}$ and we can state the main result of this section:
\begin{prop}\label{smooth}
The mapping $\Phi: \tilde{B}\times A\lra \tilde{B}$ defined in prop. \ref{ext-action}
is smooth. 
\notka{smooth}
\end{prop} 
{\em Proof:} By the results of the previous section to prove the proposition
we have to show smoothness at two kinds of points: 
$(\tilde{b},a)\in B\times A$ such that $\alpha(b,a)=0$
and $(\tilde{b},a)\in W\times A$. We begin by points of the first kind.\\
Let $\alpha(s_0,y_0;z_0,U_0,1)=0.$ Let $Y:=U^{-1}((s^2-|y|^2-\alpha)z +y)$. Notice
that $Y$ is a smooth function of $(s,y,z,U)$ and the action can be written as:
$$\Phi(s,y,z,U,1)=\left\{\begin{array}{lcr}
\left(\frac{s}{|\alpha|}\,,\,\frac{Y}{\alpha}\right) & {\rm \,for\,} &\alpha\neq 0\\
\left[\left(\mu(s,y)\,,\,\frac{Y}{s}\right)\right] & {\rm \,for\,} &\alpha=0
\end{array}\right.$$
Let $x_0:=U_0^{-1}((s_0^2-y_0^2)z_0+y_0)/s_0=\frac{Y(s_0,y_0,z_0,U_0)}{s_0}$. 
Since $|x_0|=1$ there exists $k$ such that $(x_0)_k\neq 0$ i.e. $x_0\in \tilde{U}_k$.
Let $\epsilon:=sgn((x_0)_k)$
Since $Y$ is smooth one can find some neighbourhood of $(s_0,y_0,z_0,U_0)$ such that
$sgn(Y_k)=\epsilon$ and $|Y|>\delta>0$. Because $\alpha(s_0,y_0,z_0,U_0)=0$ it follows that
the image of some neighbourhood of $(s_0,y_0,z_0,U_0)$ by $\Phi$ is contained in
$\tilde{O}_k$. Using the definition of $\varphi_k^{-1}$ we obtain:
$$\varphi_k^{-1}\Phi(s,y,z,U,1)=
\left(\epsilon\mu(s,y)\,,\,\frac{\epsilon Y}{\epsilon\alpha+|Y|}\right)$$
on some neighbourhood of $(s_0,y_0,z_0,U_0)$ and this is a smooth mapping.\\
For $\alpha(s_0,y_0;z_0,U_0,-1)=0.$ We denote by $Y:=DU^{-1}(y-(1+\alpha)z)$.
Again $Y$ is smooth and the action:
$$\Phi(s,y,z,U,1)=\left\{\begin{array}{lcr}
\left(\frac{s}{|\alpha|}\,,\,\frac{Y}{\alpha}\right) & {\rm \,for\,} &\alpha\neq 0\\
\left[\left(-\mu(s,y)\,,\,\frac{Y}{s}\right)\right] & {\rm \,for\,} &\alpha=0
\end{array}\right.$$
Using the same arguments we obtain:
$$\varphi_k^{-1}\Phi(s,y,z,U,1)=
\left(- \epsilon\mu(s,y)\,,\,\frac{\epsilon Y}{\epsilon\alpha+|Y|}\right)$$
on some neighbourhood of $(s_0,y_0,z_0,U_0)$, where $\epsilon:=sgn((x_0)_k)$ 
and $x_0:=D U_0^{-1}(y_0-z_0)/s_0=\frac{Y(s_0,y_0,z_0,U_0)}{s_0}$.

Now we consider points in $W\times A$. Let $[(\mu_0,x_0)]=\varphi_k(\sigma_0,\xi_0)$.
We need the expression for $\alpha(s,y;z,d)$ in these coordinates. Using
definition of $\alpha$ and $\varphi_k$ we get:
$$\alpha(\sigma,\xi;z,1)=
\frac{1}{(1-|z|^2)(1-|\xi|)}[(1-|\xi|)(1+|z|^2-2\sigma^2 |z|^2)-2 (z^t\xi+|z|^2 \sigma\Gamma)]=:
\frac{L(\sigma,\xi,z,1)}{1-|\xi|},$$
$$\alpha(\sigma,\xi;z,-1)=
\frac{1}{(1-|z|^2)(1-|\xi|)}[(1-|\xi|)(1+|z|^2-2\sigma^2)-2 (z^t\xi+\sigma\Gamma)]=:
\frac{L(\sigma,\xi,z,-1)}{1-|\xi|},$$
where $\Gamma:=\sqrt{\sigma^2(1-|\xi|)^2+2|\xi|-1}$ and $|\xi|\neq 1$. Notice that
$L$ is a smooth function. Using the definition of the action we obtain:
$$\Phi(\sigma,\xi,z,U,1)=\left\{\begin{array}{lrl}
\left(\frac{1-|z|^2}{2|\sigma|z|^2+z^t\xi|}\,,\,\frac{-(1-|z|^2) Y}
{2(\sigma|z|^2+z^t\xi)}\right)\in B
& {\rm \,for\,} & |\xi|=1\,,\,z^t\xi+\sigma|z|^2\neq 0\\
\ [(\sigma\,,\,Y)\ ]\in W & {\rm \,for\,} & |\xi|=1\,,\,z^t\xi+\sigma|z|^2=0\\
\left(\frac{\sigma(1-|\xi|)+\Gamma}{|L|}\,,\,\frac{Y}{L}\right)\in B&{\rm \,for\,} & 
|\xi|\neq1\,,\,L\neq0\\
\ [(\sigma\,,\,\frac{Y}{\sigma(1-|\xi|)+\Gamma})\ ]\in W & {\rm \,for\,} & 
|\xi|\neq1\,,\,L=0\end{array}\right.,$$
where $$Y:=U^{-1}(\xi+\frac{2}{1-|z|^2}((1-|\xi)(\sigma^2-1)+z^t\xi+\sigma\Gamma)z)=
U^{-1}(\xi+\frac{2}{1-|z|^2}((1-|\xi)(2 \sigma^2-1)+2\sigma\Gamma-L)z)$$ is a smooth
function. Since for $|\xi|=1\,,\,L(\sigma,\xi,z,1)=\frac{-2(z^t\xi+|z|^2\sigma)}{1-|z|^2}$,
the action can be written as:
$$\Phi(\sigma,\xi,z,U,1)=\left\{\begin{array}{lrl}
\left(\frac{\sigma(1-|\xi|)+\Gamma}{|L|}\,,\,\frac{Y}{L}\right)\in B&{\rm \,for\,} & 
L\neq0\\
\ [(\sigma\,,\,\frac{Y}{\sigma(1-|\xi|)+\Gamma})\ ]\in W & {\rm \,for\,} & 
L=0\end{array}\right.$$

Let $z_0^t\xi_0+\sigma_0|z_0|^2\neq 0$. Since 
$L(\sigma_0,\xi_0,z_0,1)=\frac{-2(z_0^t\xi_0+\sigma_0|z_0|^2)}{1-|z_0|^2}$ we can find 
neighbourhood of this point with $L\neq 0$ on it. Then the image of this neighbourhood
by the action is contained in $B$ and the action is:
$$\Phi(\sigma,\xi,z,U,1)=
\left(\frac{\sigma(1-|\xi|)+\Gamma}{|L|}\,,\,\frac{Y}{L}\right)$$
so it is smooth.

If $z_0^t\xi_0+\sigma_0|z_0|^2=0$, let 
$x_0:=U_0^{-1}(\xi_0+2\sigma_0 z_0)=Y(\sigma_0,\xi_0,z_0,U_0)$. Let
$\epsilon:=sgn((x_0)_l)\neq 0$, i.e. $[(\sigma_0,x_0)]\in \tilde{O}_l$.
We are going to show that the image of some neighbourhood of $(\sigma_0,\xi_0,z_0,U_0,1)$
by the action is contained in $ \tilde{O}_l$.
Firstly, there exists a neighbourhood of $(\sigma_0,\xi_0,z_0,U_0)$ such that $sgn(Y_l)=\epsilon$
and $|Y|>\delta>0$ on this neighbourhood. Secondly, since 
$\sigma_0(1-|\xi_0|)+\Gamma(\sigma_0,\xi_0)=1$, there exists a neighbourhood of
$(\sigma_0,\xi_0,z_0,U_0)$ such that $sgn((\frac{Y}{\sigma(1-|\xi|)+\Gamma})_l)=\epsilon$.
Thirdly, because $L(\sigma_0,\xi_0,z_0)=0$, $\left|\frac{Y}{L}\right|>2$ on some neighbourhood of 
$(\sigma_0,\xi_0,z_0,U_0)$. So the  image of the intersection of these  neighbourhoods 
is contained in $\tilde{O}_l$. Now using the definition of $\varphi_l^{-1}$ we obtain:
$$\varphi_l^{-1}\Phi(\sigma,\xi,z,U,1)=\left(\sigma\epsilon\,,\,\frac{Y}{L+\epsilon|Y|}\right)$$
and this is a smooth mapping on some neighbourhood of $(\sigma_0,\xi_0,z_0,U_0,1)$.

It remains to show smoothness at points with $d=-1$. Proceeding exactly as in the previous case
one gets:
$$\Phi(\sigma,\xi,z,U,-1)=\left\{\begin{array}{lrl}
\left(\frac{\sigma(1-|\xi|)+\Gamma}{|L|}\,,\,\frac{Y}{L}\right)\in B&{\rm \,for\,} & 
L\neq0\\
\ [(\sigma\,,\,\frac{Y}{\sigma(1-|\xi|)+\Gamma})\ ]\in W & {\rm \,for\,} & 
L=0\end{array}\right., $$
where this time $L=L(\sigma,\xi,z,U,-1)$ and 
$$Y:=DU^{-1}\left(\xi+\frac{2}{1-|z|^2}[(1-|\xi|)(\sigma^2-1)+z^t\xi+\sigma\Gamma]z\right)=
DU^{-1}(\xi-(L+1-|\xi|)z).$$

For $z_0^t\xi_0+\sigma_0\neq0$ we obtain:
$$\Phi(\sigma,\xi,z,U,-1)=
\left(\frac{\sigma(1-|\xi|)+\Gamma}{|L|}\,,\,\frac{Y}{L}\right)\in B$$
on some neighbourhood of $(\sigma_0,\xi_0,z_0,U_0,-1)$.

For $z_0^t\xi_0+\sigma_0=0$ the action restricted to some neighbourhood is:
$$\varphi_l^{-1}\Phi(\sigma,\xi,z,U,-1)=\left(-\sigma\epsilon\,,\,\frac{Y}{L+\epsilon|Y|}\right),$$
where $\epsilon:=sgn((DU_0^{-1}\xi_0)_l)$.

It is clear that in both cases the action is smooth. The proposition is proven.\\
\dow

Let's  describe orbits of the extended action. For $m\geq 0$ let 
$W_m:=\{[(m,x)]: x\in S^{n-1}\}$ and $\tilde{H}_m:=H_m\cup H_{-m}\cup W_m$
($H_m$ was defined in (\ref{defregions})).  
\begin{prop}\label{fullorbits}
\notka{fullorbits}
 \begin{enumerate}
 \item $\tilde{H}_m$ is a submanifold of $\tilde{B}$, 
 closed for $m\neq 1$. $\tilde{H}_m$ is connected for $m <1$ 
and has two connected components for $m\geq 1$;
 \item There are the following orbits of  $A_+$: 
  \begin{itemize} 
  \item[(a)] $\{(1,0)\}$,
  \item[(b)] $H_m^+$ for $m\geq 1$, 
 \item[(c)] $\Phi_J(H_m^+)=H_m^-\cup H_{-m}\cup W_m$ for $m\geq 1$,
 \item[(d)] $\tilde{H}_m$ for $m<1 $;
  \end{itemize}
 \item There are the following orbits of $A$:
\begin{itemize}
 \item[(a)] $\{(1,0)\}$,
 \item[(b)] $\tilde{H}_m$ for $m\geq 0$. 
\end{itemize}
\end{enumerate}
\end{prop}
{\em Proof:} We will use the following:
\begin{lem} Let $\tilde{\mu}:B\ni(s,y)\mapsto |\mu(s,y)|\in R$.
\begin{enumerate}
\item The function $\tilde{\mu}$ extends uniquely
to a continous function on $\tilde{B}$;
moreover $\tilde{H}_m=\tilde{\mu}^{-1}(m)\,,\,m\neq 1$ and 
$\tilde{H}_1=\tilde{\mu}^{-1}(1)\setminus\{(1,0)\}$;
\item $\tilde{\mu}$ restricted to $\tilde{B}\setminus (\tilde{H}_0\cup \{(1,0)\})$ is a smooth 
submersion.
\end{enumerate}
\end{lem}
{\em Proof of the lemma:} 1) Uniqueness is obvious. Let 
$f: R\times R^n\ni(\sigma,\xi)\mapsto |\sigma|$ and let us define 
$\tilde{\mu}_k:\tilde{O}_k\lra R$ by:
$$\tilde{\mu}_0:=|\mu|\,\,,\,\,
\tilde{\mu}_k=f\,\cdot \tilde{\varphi}_k\,\,,\,k=1,\dots,n$$
It is clear that each of them is a continuous function, so it remains to 
show that $\tilde{\mu}_k=\tilde{\mu}_l$ on $\tilde{O}_k\cap\tilde{O}_l$, 
but this is obvious from the definition of $\tilde{\varphi}_k$.
The second part of this  statement is clear.\\
2) The local representatives of $\tilde{\mu}$ are $|\mu|$ on $B$ and $f$ on 
$\tilde{O}_k\,,\,k\geq 1$. It is clear that they are smooth out of the set
$\mu^{-1}(0)$ and $f^{-1}(0)$ i.e. out of $\tilde{H}_0$. Moreover $\tilde{\mu}'$ has 
rank $1$ out of the point $(1,0)$.\\\dowl\\
1. Now it is clear that $\tilde{H}_m$ is closed except for $m=1$. Using the 
expression for the action ($\ref{ext-1}$)  one checks that for $m\geq 1$ 
$\phi_J(H_m^+)=H_m^-\cup H_{-m}\cup W_m$. Therefore 
$\tilde{H}_m=H_m^+\cup\phi_J(H_m^+)$, since each of the components is 
connected and they are disjoint $\tilde{H}_m$ has two connected components for
$m\geq 1$. For $m>0$ let $S_m:=\{(s,y): s=\frac1m=|y|\}$. Again using ($\ref{ext-1}$)
one obtains $W_m=\phi_J(S_m)$, so $W_m$ is connected. For $1>m>0$, $\overline{H_m}=H_m\cup W_m$ and
$\overline{H_{-m}}=H_{-m}\cup W_m$, since $H_m$ and $H_{-m}$ are connected $\tilde{H}_m$ is 
connected. For $m=0$, $\tilde{H}_0=\overline{H_0}$ and since $H_0$ is connected,
$\tilde{H_0}$ is connected. \\
That $\tilde{H}_m$ is a submanifold for $m\neq 0$ follows from the lemma. For $m=0$ it is clear that
$H_0$ is a submanifold of $B$, therefore also in $\tilde{B}$. Moreover one immediately
checks that for $w\in W_0$ the maps $\tilde{\varphi}_k$ defined in lemma \ref{atlas} have the
submanifold property.\\
2. and 3. These are direct consequences of the prop. \ref{orbits} and the definition of the action
(\ref{ext+1},\ref{ext-1}).
\\\dow
\section{Global structure of the manifold $\tilde{B}$}\label{section-global}

\begin{prop}\label{orient}
\notka{orient}
Let  $n\geq 2$. The manifold $\tilde{B}$ is orientable if and only if $n$ is even.
\end{prop}
{\em Proof: } Let $n$ be even. Consider the atlas on $\tilde{B}$ defined in 
the previous section. We have: 
$$\varphi_k^{-1}\varphi_l(\sigma,\xi)=\left\{\begin{array}{lcr} 
(\sigma\,,\, \xi) & {\rm for} &(\sigma,\xi)\in D_{++}\\
(-\sigma\,,\, \frac{\xi}{1-2 |\xi|}) & {\rm for} &(\sigma,\xi)\in D_{+-}
\end{array}\right.,$$
where $D_{++}, D_{+-}$ are disjoint open sets in $R\times R^n$. 
The derivative of this mapping on $D_{+-}$
is equal:
\notka{fik-fil-der}
\begin{equation}\label{fik-fil-der} 
\left(\begin{array}{cc} -1 & 0\\0 & \frac{1}{(1-2|\xi|)^2}\left((1-2|\xi|)I+\frac{2}{|\xi|}\xi\xi^t\right)
\end{array}\right)
\end{equation}
and its determinant: $-(1-2|\xi|)^{-1-n}>0$ since $n$ is even and $(1-2|\xi|)<0$ on $D_{+-}$.

For $k=0$ we have:
$$\varphi_0^{-1}\varphi_l(\sigma,\xi)=\left(\frac{\sigma(1-|\xi|)+\Gamma}{|1-|\xi||}\,,\,
\frac{\xi}{1-|\xi|}\right)\,\,,\,
|\xi|\neq 1\,\,,\,\Gamma:=\sqrt{\sigma^2(1-|\xi|)^2+2|\xi|-1}$$
and derivative:
\notka{fik-der}
\begin{equation}\label{fik-der}
\left(\begin{array}{cc} sgn(1-|\xi|)\left(1+\frac{\sigma(1-|\xi|)}{\Gamma}\right) & 
\frac{sgn(1-|\xi|)}{(1-|\xi|)^2\Gamma}\xi^t\\
0 & \frac{1}{|\xi|(1-|\xi|)^2}\left(|\xi|(1-|\xi|)I+\xi\xi^t\right)\end{array}\right)
\end{equation}
The determinant is equal to: $sgn(1-|\xi|)(1+\frac{\sigma(1-|\xi|)}{\Gamma})(1-|\xi|)^{-n-1}>0$, 
so this atlas defines an orientation on $\tilde{B}$.

Let us now consider an odd $n$. Consider the action of a boost. Its derivative was computed in 
lemma \ref{der}. The determinant of the derivative is negative for $\alpha(s,y,z,1)<0$.
Since any boost can be joined  by a continuous curve with the group identity this proves that 
$\tilde{B}$ is  non-orientable for an odd $n$.\\\dow 
\begin{prop}\label{fundamental}
\notka{fundamental}
For $n>2$ the fundamental group of $\tilde{B}$  is $Z_2$.
\end{prop}
{\em Proof:} It is clear that $\tilde{B}$ is path connected. We will
use Seifert-van Kampen theorem. Let us define $X_1:=B$ and 
$X_2:=\{(s,y)\in B : |y|>2\}\cup W$. 
Then $X_1\cap X_2=\{(s,y)\in B : |y|>2\}$, $X_1, X_2, X_1\cap X_2$ are path connected and
$\pi_1(X_1)=\{e\}=\pi_1(X_1\cap X_2)$. We will use the following lemma (notation was introduced 
in the beginning of previous section).
\begin{lem} Let $\gamma:R\times [0,1]\ni(\sigma,t)\mapsto \gamma(\sigma,t)\in R$ be
continuous and $\gamma(-\sigma,t)=-\gamma(\sigma,t)$. For $k=1,\dots,n$ define the mapping:
$h_k: O_k\times [0,1]\ni(\sigma,\xi,t)\mapsto 
(\gamma(\sigma,t), \frac{\xi}{1+t(|\xi|-1)})\in R\times R^n$. Then:
\begin{enumerate}
\item $h_k(O_k\times[0,1])\subset O_k$;
\item For  $\tilde{h}_k: \tilde{O}_k\times[0,1]\lra \tilde{O}_k$ defined as 
$\tilde{h}_k:=\varphi_k\cdot h_k\cdot (\tilde{\varphi}_k\times id)$ we have 
$\tilde{\varphi}_k=\tilde{\varphi}_l$ on $\tilde{O}_k\cap\tilde{O}_l$ $k,l=1,\dots,n$;
\item There exists unique mapping 
$\tilde{h}:\bigcup_{k=1}^n \tilde{O}_k\times[0,1]\lra \bigcup_{k=1}^n \tilde{O}_k$ such that 
$\tilde{h}=\tilde{h}_k$ on $\tilde{O}_k\times[0,1]$.
\end{enumerate}
\end{lem}
{\em Proof of the lemma:} 1. This a simple consequence of definitions of $O_k$ and $h_k$.\\
2. Because of 1. $\tilde{h}_k$'s are well defined. Again this is a matter of simple 
computations (use the lemma \ref{atlas}).\\
3. This is an immediate consequence of 2.\\
\dowl

From Seifert-van Kampen theorem we obtain $\pi_1(\tilde{B})=\pi_1(X_2)$. Now take  
$\gamma(\sigma,t):=\sigma (1-t)$. For this choice  it is easy to see 
that $\tilde{h}$ can be restricted 
to $X_2$ i.e. $\tilde{h}(X_2\times [0,1])\subset X_2$, moreover :
$$\tilde{h}(x,0)=x\,,\,\tilde{h}(x,1)\in W_0\,\,{\rm for }\,\,x\in X_2\,\,
{\rm and}\,\, \tilde{h}(w,t)=w\,\,{\rm for}\,\, w\in W_0,$$ 
so $W_0$ is a strong deformation retract of $X_2$. Since $W_0=RP^{n-1}$ and 
$\pi_1(RP^n)=Z_2$ for $n\geq 2$, the proposition is proven.\\\dow

Now we are going to show that $\tilde{B}$ can be given a structure of a Lorentzian manifold.\\
Consider the metric $\tilde{g}$ on $B$ given in 
coordinates $(s,y)$ by the matrix $g(s,y):=\frac{1}{s^2}\eta$,
where $\eta$ is the standard Lorentz metric on $R\times R^n$.
\begin{prop}\label{metric}
\notka{metric}
\begin{enumerate}
\item $\tilde{g}$ extends uniquely to a metric on $\tilde{B}$;
\item $\tilde{g}$ is $A$-invariant.
\item $(\tilde{B},\tilde{g})$ has a constant curvature.
\end{enumerate}
\end{prop}
{\em Proof:} 1. Since $B$ is dense in $\tilde{B}$ the uniqueness is 
obvious. So we have to prove existence. For $k=1,\dots,n$ let 
$g_k(\sigma,\xi)$ be the matrix of $\tilde{g}$ in coordinates $(\sigma,\xi)$.
Using the formula for $\varphi_k$ (\ref{def-fik}) and its derivative (\ref{fik-der})
one gets:
\renewcommand{\arraystretch}{1.5}
$$g_k(\sigma,\xi):=\left(\begin{array}{cc}
\frac{(1-|\xi|)^2}{\Gamma^2} & \frac{1}{(\sigma(1-|\xi|)+\Gamma)^2}\xi^t\\
\frac{1}{(\sigma(1-|\xi|)+\Gamma)^2}\xi & \frac{1}{(\sigma(1-|\xi|)+\Gamma)^2}
\left[ -I+\frac{2+\sigma^2(|\xi|-2)}{|\xi|\Gamma^2}\xi \xi^t\right]
\end{array}\right)\,,\,\,\xi_k>0\,,\,\,\frac{1}{2}<|\xi|<2\,,\,\,|\xi|\neq 1,$$
where $\Gamma=\sqrt{\sigma^2(1-|\xi|)^2+2|\xi|-1}$. \\
Since matrix elements are smooth also for $|\xi|=1$ this expression gives $g_k$ on
the whole set $\tilde{O}_k$. Using (\ref{fik-fil-der}) one easily checks that really $\tilde{g}$
is well defined i.e. $g_k$'s glue together correctly. It remains to show 
that $\tilde{g}$ is in fact metric i.e. we have to check
its signature on $W$ (i.e. for $|\xi|=1$). \\
For $|\xi|=1$ we have:
\renewcommand{\arraystretch}{1.2}
$g_k(\sigma,\xi)=\left(\begin{array}{cc}
0 & \xi^t\\
\xi & -I+(2-\sigma^2)\xi \xi^t
\end{array}\right).$
Pick a basis $e_0:=\left(\begin{array}{c}\rho\\\xi\end{array}\right)\,$,
$\,e_1:=\left(\begin{array}{c}-\frac{1}{\rho}\\\xi\end{array}\right)\, $,
$\,e_l:=\left(\begin{array}{c}0\\f_l\end{array}\right)\,,\,l=2,\dots,n\,,\,\xi^t f_l=0$, where
$\rho:=\frac{1}{2}(\sigma^2-1+\sqrt{4+(1-\sigma^2)^2})$. Short calculation shows that in this
basis 
$$g_k(\sigma,\xi)=\left(\begin{array}{ccc}
\sqrt{4+(1-\sigma^2)^2} & 0 & 0\\
0 & -\sqrt{4+(1-\sigma^2)^2} & 0\\
0 & 0 & -I
\end{array}\right).$$
So $g_k$ is nondegenerate and has the correct signature.\vs\\
2. Using formulae for derivative of the action computed in lemma \ref{der} one checks that 
$\phi^*_a(\tilde{g}(s,y))=\tilde{g}(\phi_a(s,y))\,,\,{\rm for\,\,} a\in A\,\,,\,\,(s,y)\in D_a$
i.e. $\phi^*_a(\tilde{g})-\tilde{g}=0$ on $D_a$. But since $D_a$ is dense in $\tilde{B}$ and
the left hand side is a smooth section of a vector bundle it must be $0$ everywhere.\\
3. Define the coordinates on $B$ : $(E,y):=(-\log s, y)$. In this coordinates 
$\tilde{g}=dE^2-e^{2E}(dy_1^2+\dots dy_n^2)$, so locally $B$ is the de Sitter space. 
So $\tilde{g}$ has a constant curvature on $B$, therefore on $\tilde{B}$.\\\dow\\

Finally let us end this paper by giving a  more clear description of $\tilde{B}$ and the action of $A$. 
To this end we use appriopriate extension of the mapping  (\ref{intertwiner}). 
Let $B_1:=\{(E,P)\in R\times R^n: E^2-|P|^2>-1\}$, $W_1:=\{(E,P)\in R\times R^n: E^2-|P|^2=-1\}/\sim$, where
$(E,P)\sim (-E,-P)$, and $\tilde{B}_1:=B_1\cup W_1$ (i.e.  $\tilde{B}_1$ is the interior of one-sheeted hyperboloid, 
together with the hyperboloid with opposite points identified, of course which hypeboloid we choose is irrelevant).
Let the action of $A\ni a=(z,U,d)$ on $R^{n+1}$ be defined by:
$$\phi_0(a)(E,P)=\phi_0(z,U,d)(E,P):=\left(\begin{array}{cc}
\frac{1+|z|^2}{1-|z|^2} & \frac{2}{1-|z|^2}z^tUD\\ 
\frac{2 }{1-|z|^2} z & (I+\frac{2}{1-|z|^2}z z^t)UD\end{array}\right) \left(\begin{array}{c} d \,E \\ d \,P\end{array}\right)$$
i.e. the proper Lorentz group acts in the standard way but the action of reflections is modified by the total inversion 
$(E,P)\mapsto (-E, -P)$. Then $\Phi_0(a,E,P):=\phi_0(a^{-1})(E,P)$ is a right action and it is clear that it 
defines the action on  $\tilde{B}_1$. Let us also define the mapping $\Psi: \tilde{B}\lra \tilde{B}_1$ by:
$$B\ni (s,y)\mapsto \left\{\begin{array}{cr} 
\left(\frac{sgn(\mu) (1-\mu s)}{s \sqrt{1+|\mu|}}, \frac{sgn(\mu) y}{s \sqrt{1+|\mu|}}\right)\in B_1 & \mu\neq 0\\
\left[\left(\frac1s,\frac{y}{s}\right)\right]\in W_1 & \mu=0\end{array}\right.,$$
$$W \ni [(\mu,x)]\mapsto \left\{\begin{array}{cr}
\left(\frac{-|\mu|}{\sqrt{1+|\mu|}},\frac{sgn(\mu) x}{s\sqrt{1+|\mu|}}\right)\in B_1 & \mu\neq 0\\
\left[\left(0, x\right)\right]\in W_1  & \mu=0 \end{array}\right.$$ 
The proof of the following proposition is straightforward but rather tedious.
\begin{prop}
$\Psi$ is a diffeomorphism and it intertwines the actions $\Phi$ and $\Phi_0$ i.e.
$\Psi \Phi=\Phi_0(\Psi\times id)$.
\end{prop}
\dowl
\section{Appendix--solution of the equation (\ref{basic-eq})}

We get the following system of equations for matrix element:\\
$(1,1)$:
$$\left(\frac{s^2+1+|y|^2}{2 s}\right)
\left(\frac{1+|z|^2}{1-|z|^2}\right)-\frac{2 y^t z}{1-|z|^2}=
\left(\frac{1+|\tilde{z}|^2}{1-|\tilde{z}|^2}\right)
\left(\frac{\tilde{s}^2+1+|\tilde{y}|^2}{2 \tilde{s}}\right)-\frac{2
}{1-|\tilde{z}|^2}\tilde{z}^t\tilde{U}\tilde{D}\tilde{y}$$
$(1,2)$:
$$\left(\frac{s^2+1+|y|^2}{2 s}\right)\left(\frac{2}{1-|z|^2}\right)z^t U D-
\frac{1}{s}y^t\left(I+\frac{2}{1-|z|^2}z z^t\right)U D=
\left(\frac{\tilde{s}^2+1+|\tilde{y}|^2}{2 \tilde{s}}\right)
\left(\frac{1+|\tilde{z}|^2}{1-|\tilde{z}|^2}\right)+
\frac{2}{1-|\tilde{z}|^2}\tilde{z}^t \tilde{U}\tilde{D}$$
$(1,3)$:
$$\frac{s^2-1+|y|^2}{2 s}\, d=
\left(\frac{1+|\tilde{z}|^2}{1-|\tilde{z}|^2}\right)
\left(\frac{\tilde{s}^2-1+|\tilde{y}|^2}{2 \tilde{s}} \right)-
\frac{2}{1-|\tilde{z}|^2}\tilde{z}^t\tilde{U}\tilde{D}\tilde{y}$$
$(2,1)$:
$$- \frac{1+|z|^2}{1-|z|^2}\, y+\frac{2}{1-|z|^2}\, z=
\left(\frac{2}{1-|\tilde{z}|^2}\right)
\left(\frac{\tilde{s}^2+1+|\tilde{y}|^2}{2 \tilde{s}}\right) \tilde{z}-
\left(I+\frac{2}{1-|\tilde{z}|^2}\tilde{z}
  \tilde{z}^t\right)\tilde{U}\tilde{D}\tilde{y}$$
$(2,2)$:
$$-\frac{2}{1-|z|^2}\,y z^t U D+\left(I+\frac{2}{1-|z|^2}z z^t\right)U D=
-\frac{2}{\tilde{s}(1-|\tilde{z}|^2)}\,\tilde{z}\tilde{y}^t+
\left(I+\frac{2}{1-|\tilde{z}|^2}\tilde{z} \tilde{z}^t\right)\tilde{U}\tilde{D}$$
$(2,3)$:
$$- d y=\left(\frac{2}{1-|\tilde{z}|^2}\right)
\left(\frac{\tilde{s}^2-1+|\tilde{y}|^2}{2 \tilde{s}}\right)\, \tilde{z}-
\left(I+\frac{2}{1-|\tilde{z}|^2}\tilde{z} \tilde{z}^t\right)\tilde{U}\tilde{D}\tilde{y}$$
$(3,1)$:
$$\left(\frac{s^2-1-|y|^2}{2s}\right)
\left(\frac{1+|z|^2}{1-|z|^2}\right)+\frac{2}{s(1-|z|^2)}y^t z=
\frac{\tilde{s}^2-1-|\tilde{y}|^2}{2 \tilde{s}}\tilde{d}$$
$(3,2)$:
$$\left(\frac{s^2-1-|y|^2}{2s}\right)\left(\frac{2}{1-|z|^2}\right) z^t U D+
\frac{1}{s}y^t\left(I+\frac{2}{1-|z|^2}z z^t\right)U D=
\frac{\tilde{d}}{\tilde{s}}\,\tilde{y}^t$$
$(3,3)$:
$$\frac{s^2+1-|y|^2}{2 s} d=\frac{\tilde{s}^2+1-|\tilde{y}|^2}{2
  \tilde{s}}\tilde{d}$$
\noindent
{\em Solution:\,\,} 
Subtractig $(3,1)$ from $(3,3)$ we obtain:
  $\frac{\tilde{d}}{\tilde{s}}=\frac{\alpha}{s}$; so  $\alpha\neq 0$,
  $\tilde{d}=sgn(\alpha)$ and $\tilde{s}=\frac{s}{|\alpha|}$.
Having $\frac{\tilde{d}}{\tilde{s}}$ we use  $(3,2)$ to find
  $\tilde{y}=\frac{D U^{-1}}{\alpha}(y+\frac{s^2-1-|y|^2+2 z^t y}{1-|z|^2}\,z)$.
Subtractig $(2,3)$ from $(2,1)$ we get $\tilde{z}$; then  $(2,2)$ gives  $\tilde{U}$.
And finally we check that $\tilde{d},\tilde{s},\tilde{y},\tilde{z},\tilde{U}$
  satisfy equations  $(2,1)$, $(1,1)$, $(1,2)$,  $(1,3)$.

\ed